\documentclass[11pt, a4paper]{article}    
\usepackage{amsmath,amssymb,latexsym}
\usepackage{graphicx}
\usepackage{bbm}
\usepackage{amsfonts}

\newcommand{\caD}{{\mathcal D}}

\newcommand{\caX}{{\mathcal X}}

\newcommand{\bbR}{{\mathbb R}}

\newcommand{\opunit}{\text{1}\kern-0.22em\text{l}}

\DeclareMathAlphabet{\mathpzc}{OT1}{pzc}{m}{it}

\newcommand{\mean}[1]{{\left< #1 \right>}}

\newcommand{\id}{\textrm{d}}

\begin{document}

\title{ Nonequilibrium Linear Response for Markov Dynamics,\\
 II: Inertial Dynamics}

\author{ Marco Baiesi$^{1,2}$, Eliran Boksenbojm$^2$, Christian Maes$^2$, Bram Wynants$^{2,*}$\\
1: Dipartimento di Fisica, Universit\`a di Padova, Via Marzolo 8, 35131 Padova, Italy \\
2: Instituut voor Theoretische Fysica, K.~U.~Leuven, B-3001 Leuven, Belgium\\
*{\tt bram.wynants@fys.kuleuven.be}
}

\date{\today}

\maketitle

\begin{abstract}
We continue our study of the linear response of a nonequilibrium
system. This Part~II concentrates on models of open and driven
inertial dynamics but the structure and the interpretation of the
result remain unchanged: the response can be expressed as a sum of
two temporal correlations in the unperturbed system, one entropic,
the other frenetic.  The decomposition arises from the
(anti)symmetry under time-reversal on the level of the
nonequilibrium action. The response formula involves a statistical
averaging over explicitly known observables but, in contrast with
the equilibrium situation, they depend on the model dynamics in
terms of an excess in dynamical activity.  As an example, the
Einstein relation between mobility and diffusion constant is
modified by  a correlation term between the position and the
momentum of the particle.
\end{abstract}

\section{Introduction}

The linear response of equilibrium systems to a time-dependent
shift in the energy has been extensively studied
throughout the twentieth century.  It is an important
 question in the construction of nonequilibrium statistical mechanics
 to find suitable extensions, most interestingly with a systematic physical interpretation.
Many efforts have been devoted to such a programme \cite{bala,ca,diez,har,ku,lipp,lipp2,mprf,mar,pug,pug2,rue,sei},
 and the present paper is contributing
a second part to what we believe is a unifying and yet novel
approach to the subject.
  In the previous Part~I \cite{jsp}
  we have shown how to think of this extension for Markov jump processes
  and for overdamped diffusions.
   In the present continuation we deal with more general processes including
    underdamped diffusions.  Their
     state space contains both position and momentum degrees of freedom
and hence there are important changes with respect to the analysis in Part~I,
mostly having to do with the nature of time-reversal and dissipation.
 For inertial systems the time-reversal includes changing the sign of the momenta, which
are the variables connected with energy dissipation, as
this is power lost by the frictional forces exerted on momenta.
More generally, the Hamiltonian structure
   adds some nontrivial aspects to the analysis.
Nevertheless the strategy with its statistical
interpretation holds basically unchanged.

 The response formula
gives the generalized susceptibilities
 for changes in the potential $U \rightarrow U - h_s \,V$ depending on
 time $s$ in a quantity $Q$ at time $t\ge s$ in the form~\cite{prl}
\begin{equation}\label{form} \langle Q(t)\rangle^h - \langle Q(t)\rangle
= \frac 1{2}\langle {\cal S}(\omega) \,Q(t)\rangle
- \frac 1{2}\langle \tau(\omega) \,Q(t)\rangle
\end{equation}
where $\omega$ is the history, ${\cal S}(\omega)$ is the
path-dependent excess in entropy flux and $\tau(\omega)$ is the
path-dependent excess in dynamical activity to linear order in $h_s$,
called frenesy~\cite{jsp}.
  Excess refers to the perturbation $V$.
The brackets $\mean{\ldots}$ and $\mean{\ldots}^h$
denote expectations in the unperturbed and perturbed process, respectively.
The structure is thus similar to that in equilibrium, with responses
written in term of correlations estimated in the unperturbed regime.
Before giving the general explanations, we start with an example,
an inertial Brownian motion under nonequilibrium driving.

 We consider a particle with momentum and position
$(q,p)\in \bbR^{2d}$ undergoing
\[
\id q = p \,\id t
\]
\begin{equation}\label{1ste}
\id p = -\nabla U(q) \,\id t + F(q)\,\id t -\gamma p \,\id t + h_t
E \,\id t + \sqrt{2\gamma/\beta}\,\id B(t)
\end{equation}
where $\gamma$ is the friction by the heat bath at inverse temperature $\beta$
and $\id B(t)$ is a standard white noise.
With a sufficiently confining potential $U$, a
stationary distribution $\rho$ settles for a nonequilibrium
due to the presence of a non-gradient force $F$. The perturbation is
in the form of a constant external field $E$ with amplitude
$h_t$. We look at the excesses, first in entropy flux: for a path
$\omega$ during $[0,t]$,
\begin{equation}\label{12}
 {\cal S}(\omega)  =  \beta \int_{0}^{t} \id s\, E\cdot p(s)\, h_s
 \end{equation}
 is the excess entropy flux by the field $E$ (heat transmitted to the reservoir
divided by the temperature, taking Boltzmann's constant equal to one).  Secondly we must
identify the path-dependent frenesy, that is, the linear
 order (in $h$) in the excess of dynamical activity or generalized escape
 rate.  We give it here without further explanation to return to
 it in Sections \ref{II} and \ref{under}.  For the present example it is given by
\begin{equation}\label{fre}
\tau(\omega)  =  - \frac{\beta}{\gamma}
\int_{0}^{t}\,h_s\, \left[\id s \big(\nabla U - F\big)
 + \id p(s) \right]\cdot E
\end{equation}
where in the square brackets and from \eqref{1ste} one discovers the
thermostating forces for the original dynamics,
friction plus forcing from the heat bath:
\[
\big(\nabla U - F\big)\,\id s + \id p(s) = -\gamma p \,\id s +
\sqrt{2\gamma/\beta}\,\id B(s)
\]
Observe that $\tau$ is symmetric under time-reversing the paths
(together with flipping the momentum).
If for example we take the momentum $p(t)$ as the observable for which we seek the
  linear response,  the result is like \eqref{form}:
  \[
  \langle p(t) \rangle^{h} - \langle p(t) \rangle =
\frac 1{2} \langle {\cal S}\,p(t)\rangle - \frac 1{2} \langle
\tau\,p(t)\rangle \] or, in differential form, yielding the
nonequilibrium dynamic mobility with $0<s<t$,
\begin{eqnarray}\label{mobi}
{\frac{\delta}{\delta h_s}\left< p(t) \right>^{h}_\rho}\biggr\vert_{h=0} & =&
  \frac{\beta }{2} \left< E\cdot p(s)\, p(t) \right>_\rho\nonumber\\
 & & + \frac{\beta}{2\gamma}  \left< E\cdot\left[ \nabla U(q(s)) - F(q(s))
 \right]\, p(t)\right>_\rho + \frac{\beta }{2\gamma}
  \frac{\id}{\id s} \left< E\cdot p(s) \,p(t) \right>_\rho
\end{eqnarray}
We have added the subscript $\rho$ to indicate that we are looking
here at the stationary situation, although this is just a standard
choice and is not necessary for our approach.  In equilibrium,
when $F$ would itself be a gradient, one has time-reversal
invariance and one easily checks
 that the second line in \eqref{mobi} contributes exactly the same as in
  the first line, obtaining the standard relation between mobility (left-hand side)
  and diffusion
  $\beta  \left< E\cdot p(s)\, p(t) \right>^{\rm eq}$ in terms of momentum correlations.
  This example, the nonequilibrium relation between mobility and
  diffusivity, has been visited already by various groups
  including \cite{kf,kf1,blickle,ss}
  containing various and different useful response relations.
We will revisit it as Example \ref{ein}.

The purpose of the present paper is to explain the
general algorithm for identifying the terms in \eqref{form} resulting e.g. in \eqref{mobi}
 and to
add the statistical dynamical meaning.  In the next section we
explain the general strategy on the level of path-space
distributions.  Section \ref{under} fills in the details for
inertial diffusions as in the example above. In Section
\ref{inter} we connect our formulation with previous ones, in
particular those obtained via first order perturbation theory.
Finally there are more examples  and comparisons in Section
\ref{exs}, including a generalized Einstein relation \eqref{er}.

\section{General strategy to linear response}\label{II}
We start with an algorithm for arriving at the general linear
response formula \eqref{form}, outlined in~\cite{prl}.
We suppose variables $(q,p)\in \bbR^{2d}$ which trace out some
path or trajectory during a time-interval $[0,t]$. We can think of
a Hamiltonian flow to which we have added  nonequilibrium forces
by various contacts with the environment. At the initial time
$s=0$ the system is found in a nonequilibrium regime, possibly a steady state.
We are interested in some observation at time $t$ when we perturb the
Hamiltonian by adding $-h_s V$, with $ t> s> 0$.
Let us also add that what follows is greatly
independent of the details of the dynamics.

 We compare the perturbed system with the original one yielding a
probability density ${\cal P}^h(\omega)$ for paths $\omega= (q(s),p(s))$
for times $s\in [0,t]$. We assume that this probability density can be
written in terms of an action ${\cal A}$, simply as \[
{\cal P}^h(\omega) \equiv \frac{\mbox{Prob}^h[\omega]}{\mbox{Prob}[\omega]} =
e^{-{\cal A}(\omega)}
 \]
That amounts to supposing $\mbox{Prob}[\omega]=0 \Rightarrow \mbox{Prob}^h[\omega]=0$.  Thus
${\cal P}^h(\omega)$ gives the changed plausibility of trajectories $\omega$ on the
considered scale of reduced description. In the case of Markov
dynamics, ${\cal A} = \int_0^t\id s\, L_s(q(s),dq(s); p(s),dp(s))$
is a time-integral. A natural decomposition of ${\cal A}$ can be
made with the time-reversal $\theta$ which changes $\omega$ into
$\theta \omega$ with $(\theta \omega)_s = (q(t-s),-p(t-s))$.
Then, ${\cal A} =  ({\cal T} - {\cal S})/2$ with
\begin{eqnarray}\label{deco}
 {\cal S} &=& {\cal A}\theta - {\cal A},\nonumber\\
 {\cal T} &=& {\cal A}\theta + {\cal A},\nonumber\\
 {\cal P}^h(\omega) &=& e^{-{\cal T}(\omega)/2}\,e^{{\cal S}(\omega)/2}
\label{S-T}
\end{eqnarray}
The source-term of time-symmetry breaking is $\cal S$
and, under typical nonequilibrium conditions, it is exactly the
excess in physical entropy flux due to the perturbation. For
Markov dynamics that is being implemented by the condition of
local detailed balance, but it can be more generally derived from
mechanical beginnings, see \cite{mnj,hal} for derivations starting
from a Hamiltonian evolution. For constructing Markov models the
condition of local detailed balance
was emphasized in \cite{leb,kls} and it underlies what
is called the fluctuation symmetry for the variable entropy
production~\cite{m,poincare,ls}. For the general algorithm, if we
know the excess entropy flux we know $\cal S$, see \eqref{12} as
a first example.

 The time-symmetric term $\cal T$ is an excess
in the time-integrated instantaneous dynamical activity.  It was
discussed in \cite{mnw,mnw1} for overdamped diffusions and for
Markov jump processes respectively.   By looking at \eqref{deco}
we see that, of two trajectories having the same entropy flux, the
one with the lowest excess in activity is most probable. Formally,
the dynamical activity in $(q,p)$ is the rate of escape from that
phase space point. We will see it implemented in Section
\ref{under} for underdamped diffusions. In the first example
\eqref{fre} and as already mentioned there,
\[
{\cal T}(\omega) = - \int_0^t \id s \,h_s\frac{\partial
V}{\partial q}(q(s))\,{\cal F}(s) + O(h^2)
\]
for random forcing ${\cal F}(s) = \frac{1}{D} [-\gamma
\,p(s) + \sqrt{2D}\,\id B(s)/\id s]$, with $D= \gamma/\beta$.

   Clearly then, the response can be calculated from
   \[
   \langle Q(t) \rangle^h =
   \langle Q(t) \,e^{{\cal S}/2 - {\cal T}/2}\rangle
   \]
   where the right-hand side averages over the unperturbed process.
   The created entropy flux
   is dissipating the extra energy in the environment which is itself of order $h$.  Therefore,
   the excess entropy $\cal S$ is already
   linear in $h$.  The dynamical activity can be linearized to ${\cal T} = \tau + O(h^2)$
    and
 we obtain \eqref{form}:
   \begin{equation}\label{fdf}
   \langle Q(t) \rangle^h =
   \langle Q(t)\rangle + \frac 1{2}\langle {\cal S}\,Q(t)\rangle   - \frac 1{2}\langle \tau\,Q(t)\rangle
   \end{equation}
   which is the same structure of linear response as in \cite{prl,jsp}.

We can in fact also consider $t=0$ in \eqref{fdf} from which we see that
   \begin{equation}\label{no}
   \langle {\cal S} Q(0)\rangle   = \langle \tau Q(0)\rangle
    \end{equation}
Since in equilibrium there is full time-reversal invariance, we
know for the right-hand side of \eqref{no} that $\langle
\tau\,Q(t)\rangle^{\textrm{eq}}= \langle \tau\, \epsilon
Q(0)\rangle^{\textrm{eq}}$, due to time-reversal invariance of
$\tau$ ($\epsilon= \epsilon_Q$ is the sign of $Q$ upon
time-reversal). Similarly $\langle {\cal S}\,Q(t)
\rangle^{\textrm{eq}} = -\langle {\cal S}\,\epsilon
Q(0)\rangle^{\textrm{eq}}$ because ${\cal S}$ is antisymmetric.
Hence, by \eqref{no},
\[
- \langle \tau\,Q(t)\rangle^{\textrm{eq}} = \langle {\cal S}\,Q(t)\rangle^{\textrm{eq}}
\]
and we recover  from \eqref{fdf} the fluctuation-dissipation relation,
\[
 \langle Q(t) \rangle^h -  \langle Q(t) \rangle^{\textrm{eq}}
 = \langle {\cal S}\,Q(t)\rangle^{\textrm{eq}}
\]
 for equilibrium, with the standard interpretation that the linear response of a
quantity to a perturbation is given by its correlation with the
entropy that the perturbation produces.
Out of equilibrium, the second term originating from the dynamical activity $\tau$,
what we like to call the frenetic term, is not reducible to the
(first)
 entropic term except for the relation \eqref{no}.  Implementing
 it in \eqref{fdf}, we can still write that for all times $t\geq 0$
  \begin{equation}\label{fdf2}
   \langle Q(t) \rangle^h -
   \langle Q(t)\rangle = \langle {\cal S}\,\frac{Q(t) - \epsilon\,Q(0)}{2}\rangle
      - \langle \tau\,\frac{Q(t)-\epsilon\,Q(0)}{2}\rangle
   \end{equation}
   where the excesses ${\cal S}$ (in entropy flux) and $\tau$ (in
   activity) are over the time-interval $[0,t]$, linear in the amplitudes $h_s$.
   The second term in the right-hand side of \eqref{fdf2} vanishes
   in equilibrium.

\section{Markov dynamics for underdamped diffusions}\label{under}

The generic model that adds a specific dynamics to the previous
section has states $(q,p) =
(q_1,q_2,\ldots,q_n;$ $p_1,p_2,\ldots,p_n)\in \bbR^{2n}$ of positions
and momenta. We imagine a weak coupling with equilibrium
reservoirs. That is: at each $1\leq i\leq n$ are attached a
standard white noise $\id B_i(t)$ with diffusion constant $D$ and a
friction coefficient $\gamma_i$ to model a heat bath at
temperature $D/\gamma_i = T_i$:
\begin{eqnarray}
\id q_i &=& p_i\, \id t \nonumber\\
\id p_i &=&
 F_i(q)-\frac{\partial U}{\partial q_i} \id t -\gamma_i p_i \id t +
  h_t \frac{\partial V}{\partial q_i} \id t + \sqrt{2D}\,\id B_i(t)
\label{ud}
\end{eqnarray}
for a given potential $U(q)$ taking care of the coupling and
pinning of the positions. The pinning is also thought to confine
the positions to some finite volume.   Besides the possibility of
having multiple temperatures $T_i$ to drive the system away  from
equilibrium, we also add a nonconservative forcing $F_i(q)$.
Finally, the perturbation is $V(q)$ and thus adds a potential to
the unperturbed Hamiltonian $H_o= \sum_i p_i^2/2 + U(q)
\rightarrow H_o - h_s\,V(q)$ with small
 time-dependent amplitude $h_s$ for $ s\geq 0$.

 Alternatively, we can write the Fokker-Planck equation for the
 unperturbed system as
 \begin{equation}\label{fk}
 \frac{\id }{\id t}\mu_t + \nabla \cdot J_{\mu_t} =0
 \end{equation}
for $\nabla = (\nabla_q,\nabla_p)$ and for current $J_\mu = (J_\mu^q,J_\mu^p)$ with
 \begin{equation}\label{curre}
 J_\mu^q= p\mu,\qquad J_\mu^p = (F-\nabla_q U)\mu - \gamma \,p \,\mu - D\,\nabla_p\mu
 \end{equation}
to be understood with diagonal matrices $\gamma$ and $D$.  The backward generator $L$ is
\[
Lf = p\nabla_q f + (F-\nabla_qU -\gamma\,p)\,\nabla_p f + \nabla_p
D \nabla_p f
\]

Let us now apply the algorithm.  In a time-interval $[0,t]$
the time-integrated entropy flux in excess (by the perturbation)
is made from the extra heat into the various reservoirs:
 \begin{equation}\label{eq:S}
{\cal S}= \sum_{i=1}^n \frac 1{T_i} \int_0^t \,h_s \frac{\partial
V}{\partial q_i}(q(s)) \,p_i(s)\,\id s
\end{equation}
On the other hand, the excess in dynamical activity is obtained by
checking the  excess action ${\cal A}$ of the previous section and
linearizing the time-symmetric part:
 \begin{equation}\label{eq:tau}
\tau(\omega) = \frac{1}{D}\sum_i \int_0^t \,h_s
\frac{\partial V}{\partial q_i}(q(s)) \,[\{F_i(q(s)) -  \frac{\partial
U}{\partial q_i}(q(s))\}\id s - \id p_i(s)]
\end{equation}
The last stochastic integral (there is no difference here between
the It\^o and the Stratonovich conventions) can still be worked out when inserting
$\tau$ in the correlation function (second term in
\eqref{form}).  We get
\begin{eqnarray}
&& \int_0^t \id s\,h_s\, \langle \frac{\partial V}{\partial q_i}(q(s)) \,
\id p_i(s)\,Q(t)\rangle_\mu = \\
&& \int_0^t \,h_s \,\frac{\id}{\id s}\langle \frac{\partial V}{\partial
q_i}(q(s)) \, p_i(s)\,Q(t)\rangle_\mu\,\id s  - \sum_j\int_0^t
\,h_s \,\langle \frac{\partial^2 V}{\partial q_j\partial
q_i}(q(s)) \, p_j(s)\,p_i(s)\,Q(t)\rangle_\mu\,\id s\nonumber
\end{eqnarray}
for an arbitrary starting density $\mu(q,p)$. As a consequence the
total linear response function for the underdamped diffusion model
\eqref{ud} is
\begin{equation}\label{genfor}
\left.\frac{\delta}{\delta h_s}\left<Q(t)\right>^h_{\mu}
\right|_{h=0} = \sum_{i=1}^n \frac 1{2T_i} \langle \frac{\partial
V}{\partial q_i}(q(s)) \,p_i(s)\,Q(t)\rangle_\mu -\frac
1{2}\langle \tau(s)\,Q(t)\rangle_\mu
\end{equation}
with
\begin{eqnarray}\label{ttt}
&& D\,\langle \tau(s)\,Q(t)\rangle_\mu =
\sum_i\langle
\frac{\partial V}{\partial q_i}(q(s))\,[F(q(s)) -
\frac{\partial U}{\partial q_i}(q(s))]\,Q(t) \rangle_\mu\\
&&
-
\frac{\id}{\id s} \sum_i \langle \frac{\partial V}{\partial q_i}(q(s))
\, p_i(s)\,Q(t)\rangle_\mu
+
\sum_{i,j} \langle \frac{\partial^2 V}{\partial q_j\partial q_i}(q(s))
\,p_j(s)\,p_i(s)\,Q(t)\rangle_\mu\nonumber
\end{eqnarray}

Note that all the observables (within the expectations) are
explicitly known.  The formula remains valid whether $\mu = \rho$ (stationary regime) or
 otherwise (transient regime).
Some specific examples follow in Section \ref{exs}.

\section{From the Dyson expansion}\label{inter}

Linear response is first order perturbation theory and here we
assume smooth behavior. We could therefore also set up a simple
Taylor expansion, or call it time-dependent perturbation theory,
as for example in \cite{ku,mar}, also for nonequilibria. Below we
sketch the immediate problem of explicitness, but true enough, the
problem is not so much finding a linear response formula but
rather to get a useful form that also yields interesting relations
with physical concepts within fluctuation theory.

We prepare the system at time $t=0$ according to its stationary
distribution $\rho$; the perturbation $-h\,V$ is added for positive times
--- we think of the set-up in \eqref{ud}. Therefore, for times
$t\geq 0$ the
dynamics has (backward) generator (working on observables)
\[
L^h =  L + h \,\nabla_q V \cdot \nabla_p,\quad L = (F-\nabla
U-\gamma p)\cdot\nabla_p + p\cdot\nabla_q + \nabla_p D\nabla_p
\]
with $\gamma$  written as diagonal matrix.  For the change in
expectations at times $t$ with respect to what we had at time zero
\[
\langle Q(t)\rangle^h - \langle Q(t)\rangle = \int \id p\id q\,
\rho(q,p)\,\big(e^{tL^h} - e^{tL}\big)Q(q,p)
\]
we  get the linear order
\[
e^{tL^h} - e^{tL} = \int_0^t e^{(t-s)L}\,(L^h-L)\,e^{sL} \id s +
O(h^2)
\]
Or, always to leading order in $h\downarrow 0$,
\[
\frac{1}{h}\big[\langle Q(t)\rangle^h - \langle Q(t)\rangle\big]
=\int_0^t\id s\,R_{QV}(t,s)
\]
with response function
\begin{equation}\label{dys}
R_{QV}(t,s) = \int \id q \,\id p\, \rho(q,p)\,\nabla_q V \cdot \nabla_p \,e^{(t-s)L}Q(q,p)
\end{equation}
(The dependence on time $s$ and on
time $t$ is made for greater generality in case the perturbation is
time-dependent through $h_s$.)
Equation \eqref{dys} is directly useful when $\rho$ is given, as in equilibrium, but not otherwise.  With
partial integration we obtain
\begin{eqnarray}\label{dys1}
R_{QV}(t,s) &=& -\langle
\frac{\nabla_p\rho}{\rho}(q(s),p(s))\cdot  \nabla_q V(s)\,Q(t)\rangle_\rho\nonumber\\
&=& \beta\frac {\id}{\id s}\langle V(s)Q(t)\rangle -
 \beta\langle[p(s) + \frac{\nabla_p\rho}{\beta\rho}(q(s),p(s))\cdot
  \nabla_q V(q(s))\,Q(t)\rangle_\rho
\end{eqnarray}
where in the second line we have introduced the equilibrium form
(first term) at inverse temperature $\beta$ as a reference. The
term $\nabla_p\rho/\rho = \nabla_p \ln \rho$ is unknown in general
which brings the lack of explicitness.

We can still make the connection with \cite{gaw}.  We define a new
current ${\cal J}_\rho$ (with respect to $J_\rho$ of
\eqref{curre}) that remains divergenceless:
\[
{\cal J}_\rho = (J^q_\rho + \nabla_p\rho/\beta,J^p_\rho -
\nabla_q\rho/\beta),\quad \nabla \cdot {\cal J}_\rho = \nabla\cdot
J_\rho = 0
\]
so that it still satisfies the same Fokker-Planck equation
\eqref{fk} (stationary for $\mu_t=\rho$).  In terms of this new
current we get
\begin{eqnarray}\label{dys2}
R_{QV}(t,s) =  \beta\frac {\id}{\id s}\langle V(s)Q(t)\rangle_\rho
- \beta\langle\frac{{\cal J}_\rho}{\rho}(s)\cdot\nabla_q
V(s)\,Q(t)\rangle_\rho
\end{eqnarray}
which is exactly the formula for the response relation in
\cite{gaw}. Indeed, as also pointed out in \cite{ss}, the
nonequilibrium fluctuation-response relation reduces to the
equilibrium fluctuation-dissipation relation when viewed from a
``Lagrangian frame,'' moving with the local velocity of the
system.

Let us now see how the above expressions are mathematically
related to our formula \eqref{form}.
As above and as in
\cite{gaw} we restrict ourselves to having one environment
temperature $\beta^{-1}$.

The crucial term to rewrite in (\ref{genfor}) and (\ref{ttt}) is
the one with the derivative with respect to time. By definition of
the backward generator $L$, such terms can be rewritten using
\[
 \frac{\id}{\id s} \langle \frac{\partial V}{\partial
q_i}(q(s)) \, p_i(s)\,Q(t)\rangle_\rho = \frac{\id}{\id s}\langle
\frac{\partial V}{\partial q_i}(q(0)) \,
p_i(0)\,Q(t-s)\rangle_\rho = - \langle \frac{\partial V}{\partial
q_i}(q(s)) \, p_i(s)\,LQ(t)\rangle_\rho
\]
where $\rho$ is again the
stationary measure. Again, this can be rewritten using the
definition of the adjoint generator $L^*$: for smooth functions
$f,g$,
\[ \int \id p \,\id q \,\rho(q,p) f(q,p) Lg(q,p) = \int \id p\, \id q\, \rho(q,p) g(q,p) L^*f(q,p) \]
 or,
\[ L^* = -p\nabla_q  - (F-\nabla_qU -\gamma\,p)\,\nabla_p  + \nabla_p
D \nabla_p  +2\frac{\nabla_p\rho}{\rho}\cdot\nabla_p  \]
Clearly $L^*$ directly depends on the stationary measure.
Its definition also implies that
\[ \langle \frac{\partial V}{\partial
q_i}(q(s)) \, p_i(s)\,LQ(t)\rangle_\rho = \langle
L^*\left(\frac{\partial V}{\partial q_i} \,
p_i\right)(q(s),p(s))\,Q(t)\rangle_\rho \] If we now use the
explicit form of $L^*$ in this, and plug the result back in
(\ref{genfor}), then we indeed get (\ref{dys1}).

The derivation above  is generalizable to the transient case, in which the
response function gets written as
\begin{eqnarray}\label{dys3}
R_{QV}(t,s) &=& -\langle
\frac{\nabla_p\mu_s}{\mu_s}(q(s),p(s))\cdot \nabla_q V(s)\,Q(t)\rangle_\mu\nonumber\\
&=& \beta\frac {\id}{\id s}\langle V(s)Q(t)\rangle_\mu
- \beta\langle\frac{{\cal J}_{\mu_s}}{\mu_s}(s)\cdot\nabla_q
V(s)\,Q(t)\rangle_\mu
\end{eqnarray}

\section{Examples}\label{exs}

In order to illustrate the algorithm in concrete terms, we collect
here a number of physically interesting examples of nonequilibrium
systems following the Langevin equations \eqref{ud}: we remind
that a potential $U$ determines conservative forces $(\partial
U/\partial q_i)$, but the degrees of freedom $(q_i,p_i)$ can be at
different temperatures  $T_i$, generating a regime out of
equilibrium. For simplicity we have fixed a diffusion constant
$D$.  The respective reservoirs are at equilibrium and they
 impose the Einstein relation with the friction coefficients $\gamma_i = T_i/D$.
Another way of going out of equilibrium is to introduce
nonconservative forces $(F_i)$, such as external fields that are
rotational. A final possibility is to start from an initial
condition that is not stationary, and thus to observe the response
in a transient regime. Numerical results are better presented with
integrated responses, for example by considering the generalized
susceptibility
\begin{equation}
  \chi(t) = \int_0^t \id s\, R_{QV}(t,s) = \frac{\mean{Q(t)}^h-\mean{Q(t)}}{h}
\end{equation}
for small constant $h_s=h$ for all times $s>0$.

\subsection{Langevin particle in a periodic potential}
Recently there have been experiments testing the response of an
overdamped particle (high viscosity limit) in a periodic
potential~\cite{cili}. In the previous paper~\cite{jsp} we have
discussed simulations of that system. Here we look for the changes in an
underdamped set-up, allowing e.g. for the particle to have a
considerable inertia and to obey a noisy Hamiltonian dynamics.

 We denote the position by $q(t) \in S^1$ (on the circle) and the momentum by $p(t)\in \bbR$
 and we choose for simplicity a particle with mass equal to
one. Then, the equations \eqref{ud} simplify to
\begin{eqnarray}\label{lang}
  \id q(t) &=& p(t) \id t\nonumber\\
  \id p(t) &=& a(t) \id t - \gamma p(t) \id t \,{ - h_t\, g(t)} \id t + \sqrt{2 D} \,\id B(t)
\end{eqnarray}
where we abbreviate
\begin{eqnarray}
a(t)  &=& f - \frac{\id U}{\id q}(q(t)) \qquad\text{(deterministic force)}\nonumber\\
{ g(t)} &=& { - \frac{\id V}{\id q}}(q(t)) \qquad\text{(perturbing force)}
\label{eq:g}\nonumber
\end{eqnarray}
The nonconservative force $f$ is the driving and is taken constant over the circle, thus
effectively tilting the conservative potential.
At time $s=0$ the unperturbed system is in the stationary nonequilibrium $\rho$
corresponding to \eqref{lang}.
Hence,  the integrated correlations are
\begin{eqnarray}
  C(t) &=& \frac{1}{h}\mean{{\cal S}(\omega) Q(t)}     \qquad \qquad\text{(entropic term)}\\
&=&- \beta \int_0^t \id s \,\mean{p(s) g(s) Q(t)}_\rho
 \nonumber\\
  K(t) &=& -\frac{1}{h}\mean{\tau(\omega) Q(t)}  \qquad \qquad\text{(minus frenetic term)}\\
 &=& \frac{1}{D} \left\{ \int_0^t ds \mean{ a(s) g(s) Q(t) }_\rho -
 \int_{0}^{t} \mean{\id p(s)  g(s) Q(t) }_\rho\right\}\nonumber\\
 C_{\textrm{ne}}(t)&=& \frac{C(t) + K(t)}{2}\label{cnt}
\end{eqnarray}
Contrary to Part~I, here we embed $\beta$'s in the definitions of
correlation functions: the integrated response relation is thus
$\chi(t) = C_{\textrm{ne}}(t)$. We take $V(q) = U(q) = \cos q$
like in previous works~\cite{cili,jsp}, and again also $Q = U$.
Since the perturbation $g$ is the gradient of a potential $V$,
\[
C^*(t) = \beta \,[ \mean{V(t)Q(t)}_\rho - \mean{V(0)Q(t)}_\rho ]
\]
is an alternative  to $C(t)$ for expressing the entropic term
(from time-integrating the excess in entropy flux).

\begin{figure}[!bt]
\begin{center}
\includegraphics[angle=0,width=13.cm]{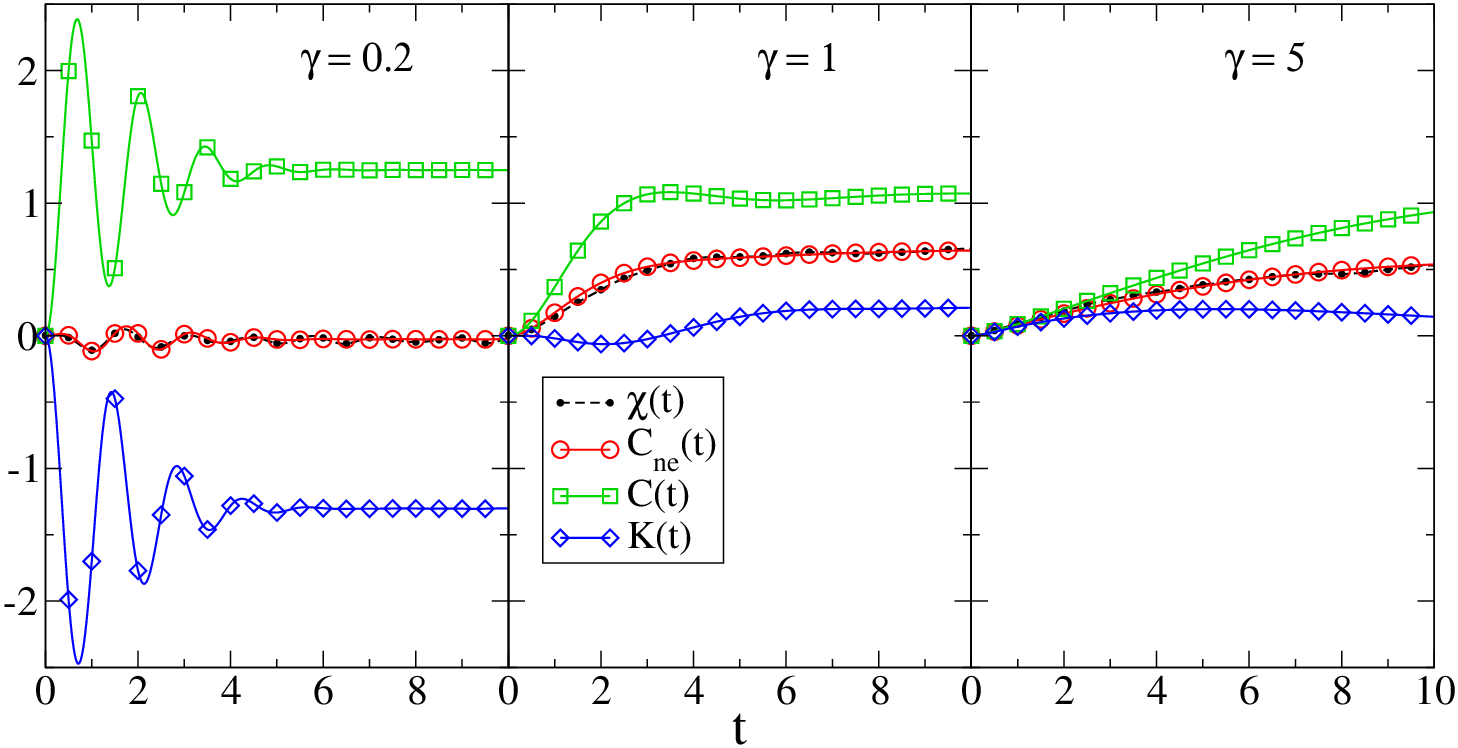}
\end{center}
\caption{Integrated correlation functions of the excess in entropy production
$C(t)$, of the frenesy $K(t)$, and their average $C_{\textrm{ne}}(t)$ giving the response in nonequilibrium,
and the integrated response $\chi(t)$ calculated directly with $h=0.01$.
Panels are for simulations with various friction coefficients:
$\gamma=0.2$ (left), $\gamma=1$ (center), and $\gamma=5$ (right).
Other parameters: $T=1/\beta=0.4$, $f=0.9$.
The inertial regime is less sensitive to the perturbation,
as $\chi(t)$ displays only a small wiggling: a high entropy
production is almost compensated by a high activity.
For larger friction we recover  previously studied overdamped scenarios;
with this setting ($f\lesssim 1$) the frenesy is close to zero compared with the excess entropy,
and their combination yields $C_{\textrm{ne}}\approx C/2$.
\label{fig:Lang}}
\end{figure}

In Fig.~\ref{fig:Lang} we visualize the various terms (see caption of the figure for more details),
for three scenarios with different viscosity, increasing from left to right.
The response is well reproduced by $C_{\textrm{ne}}(t)$ of \eqref{cnt}, even if we
perform a numerical integration with $dt=10^{-3}$.
Oscillations in the response are visible for small viscosity; at higher friction there is
 a monotonous drift toward a new stationary state
(right panel in Fig.~\ref{fig:Lang}).

\begin{figure}[!bt]
\begin{center}
\includegraphics[angle=0,width=13.cm]{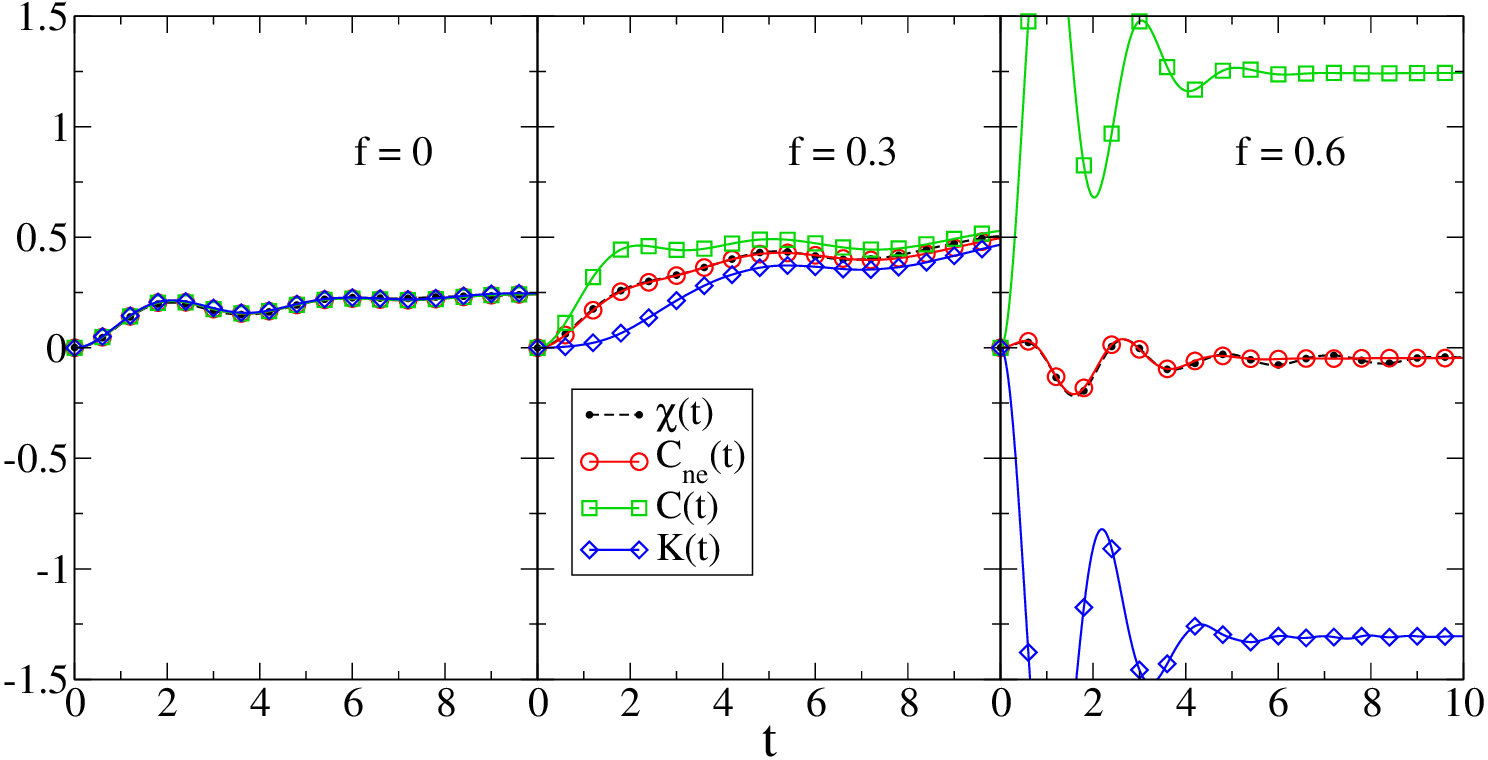}
\end{center}
\caption{As in figure \ref{fig:Lang}, but with fixed $\gamma=0.2$ and
varying force $f$, from left to right: $f=0$ (equilibrium), $f=0.3$, and $f=0.6$
(case $f=0.9$ is in the previous figure).
In equilibrium all the terms coincide, as expected.
\label{fig:Lang2}}
\end{figure}

In Fig.~\ref{fig:Lang2} we can follow the response as a function of the driving $f$, with
$f=0$ for equilibrium.  For $f\neq 0$ the entropic term $C(t)$ can be quite different from
$C_{\textrm{ne}}(t)$.

\subsection{A generalized Einstein relation}\label{ein}

We try to imagine here a collection of independent particles with
no external potential ($U=0$) except for keeping the particles in
a large cylinder filled with some fluid.  The motion is inertial
and driven, e.g. with a driving ($F(q)$) resulting from steady
differential rotation of the fluid.  We consider therefore the
general Markov dynamics (\ref{ud}), but the equation now refers to
a single particle with $j,k$ being possibly different space
directions. We are interested in the (long) transient regime in
which the position diffuses.  Initially the particle is found at
$(q(0),p(0))$ from a density $\mu$
on phase space.\\
As perturbation we take $V(q) = q_j$ for a fixed direction $j$,
and the field $h_t=h$ is supposed constant in time. To observe is
the variable  $Q = p_k$ for possibly another direction $k$. Again,
in this case the response function is related to the mobility of
the system, i.e., the way in which the average velocity changes
when a constant force is added.\\
 We fix a large time $u$ and we
define the time-averaged mobility by
\[
\caX_{jk}(u) = \frac 1{u}\int_0^u\id
t\,\left.\frac{\partial}{\partial h}\left<p_k(t)\right>^h_{\mu}
\right|_{h=0}
\]
starting from $\mu$ at time zero. We wish to connect this to the
velocity fluctuations in the unperturbed (but driven) system:
\[
\caD_{jk}(u) = \frac{1}{2u}\left<[q_j(u)-q_j(0)]\,[q_k(u)
-q_k(0)]\right>_{\mu} = \frac{1}{2u}\int_0^u \id t\int_0^u \id s
\left<p_j(s)\,p_k(t)\right>_{\mu}
\]
  We do not need to take the limit
$u\uparrow +\infty$. In equilibrium, when the Maxwell distribution
is installed with $\mu$, for $F$ and $U$ equal to zero and with
all temperatures equal to $\beta^{-1}$, then $\caX_{jk} =
\frac{1}{\gamma_j}\delta_{j,k}$ and $\caD_{jk} =
\frac{1}{\beta\gamma_j}\delta_{j,k}$. Indeed, in general
equilibrium systems the Einstein relation $\caX_{jk} =
\beta\caD_{jk}$ coincides with the fluctuation-dissipation
relation, \cite{bala}. This is no longer true for a system out of
equilibrium, see also \cite{blickle}. We give the explicit
modification.\\

We integrate the response formula (\ref{genfor})--\eqref{ttt} for
$s\in [0,u]$ where $u>t$, and then integrate $t$ again between
$[0,u]$ to obtain
\begin{eqnarray}\label{er}
  \caX_{jk}(u) &=& \beta_j \caD_{jk}(u) - \frac{1}{2Du}\int_0^u \id t\int_0^u \id s
\left< [F_j(q(s))-\frac{\partial U}{\partial q_j}(q(s))]\,p_k(t)\right>_{\mu} \\
&&+\frac{1}{2Du}\int_0^u \id t\left<
[p_j(u)-p_j(0)]\,p_k(t)\right>_{\mu}\label{ers}
\end{eqnarray}
This relation reduces
 to the familiar Einstein relation if the unperturbed system is in full
 equilibrium. For nonequilibrium, the last term \eqref{ers} is purely inertial. We can assume
exponential decay in the momentum-momentum correlation, after
which the integral remains bounded for arbitrary large $u$; but
then \eqref{ers} is arbitrary small by the prefactor $1/u$. For
the second term in \eqref{er} we can do the integral of $p_k(t)$
to insert $q_k(u) - q_k(0)$. The nonequilibrium Einstein relation
thus essentially corrects the equilibrium relation by the
correlation between the position-dependent force imposed to the
particle and the total change in position.

\subsection{Coupled oscillators}

We now consider coupled one-dimensional oscillators out of equilibrium.
First we consider the stationary regime (with density of states $\rho$)
due to different temperatures $T_i$.
 For a linear chain we can take \eqref{ud} and put an oscillator at each
$i=1,\ldots,n$. The conservative potential $U$
is the sum $U=\sum_{i=0}^n \varphi(q_{i+1}-q_i)$
of local couplings between the oscillators $i$ and $i+1$,
with $\varphi(x) = \frac{1}{2} x^2 + \frac{1}{4} x^4$.
Boundary conditions are imposed by keeping $q_0 = q_{n+1} = 0$.
A basic perturbation  is given by switching on an external
field on the $j$-th particle $V(q) = - E q_j$.
Taking the momentum $Q = p_k$ at site $k$ to be the observable,
the variable excess in entropy flux \eqref{eq:S} reduces, similarly to \eqref{12}, to
\begin{equation*}
 {\cal S}(\omega) = - \frac{E}{T_j} \int\limits_{0}^{t} \id s \:  p_{j}(s) \,h_s
\end{equation*}
while the frenesy equals
\begin{equation*}
{\tau}(\omega) = - \frac{E}{D} \int\limits_{0}^{t}  \id s
\left[ \dot{p_j}(s) + \frac{\partial U}{\partial q_j}(q(s)) \right]\, h_s
\end{equation*}

From \eqref{genfor} and \eqref{ttt} and quite similar to \eqref{mobi}, we have
\begin{equation}
\label{FDR}
\left.\frac{\delta}{\delta
h_s}\left<p_k(t)\right>^h_{\rho} \right|_{h=0}
 =
  - \frac{E}{2T_j} \left< p_{j}(s) p_{k}(t) \right>_{\rho} -
   \frac{E}{2D} \frac{\id}{\id s} \left< p_{j}(s) p_{k}(t) \right>_{\rho} -
    \frac{E}{2D} \left< \frac{\partial U}{\partial q_{j}}(q(s))\, p_{k}(t) \right>_{\rho}
\end{equation}
This last relation is still valid for all times $s,t$ and is
automatically equal to zero for $s>t$ (causality).  Let us write $\beta_j = \frac{\gamma_j}{D} = 1/T_j$ and rearrange
formula (\ref{FDR}) for the situation where $s<t$:
\begin{eqnarray*}
\left.\frac{\delta}{\delta h_s}\left<p_k(t)\right>^h_{\rho}
\right|_{h=0}
 & =
  & - E \left( \frac{\beta_j + \beta_k}{2} \right) \left< p_{j}(s) p_{k}(t) \right>_{\rho}\\
& & - \frac{E}{2 D} \left( \left< \frac{\partial
U}{\partial q_{j}}(q(s))\, p_{k}(t)\right>_{\rho} + \left<
p_{j}(s) \frac{\partial U}{\partial q_{k}} q(t) \right>_{\rho}
\right)
\end{eqnarray*}
Note that the right-hand now shows a formal space-time symmetry
for exchanging $j\leftrightarrow k, s \leftrightarrow t$. In
equilibrium the symmetry is true on spatial level alone,
$j\leftrightarrow k$, because time-symmetry is automatic.  That is
then an instance of Onsager reciprocity.\\

\begin{figure}[!tb]
\begin{center}
\includegraphics[angle=0,width=11.5cm]{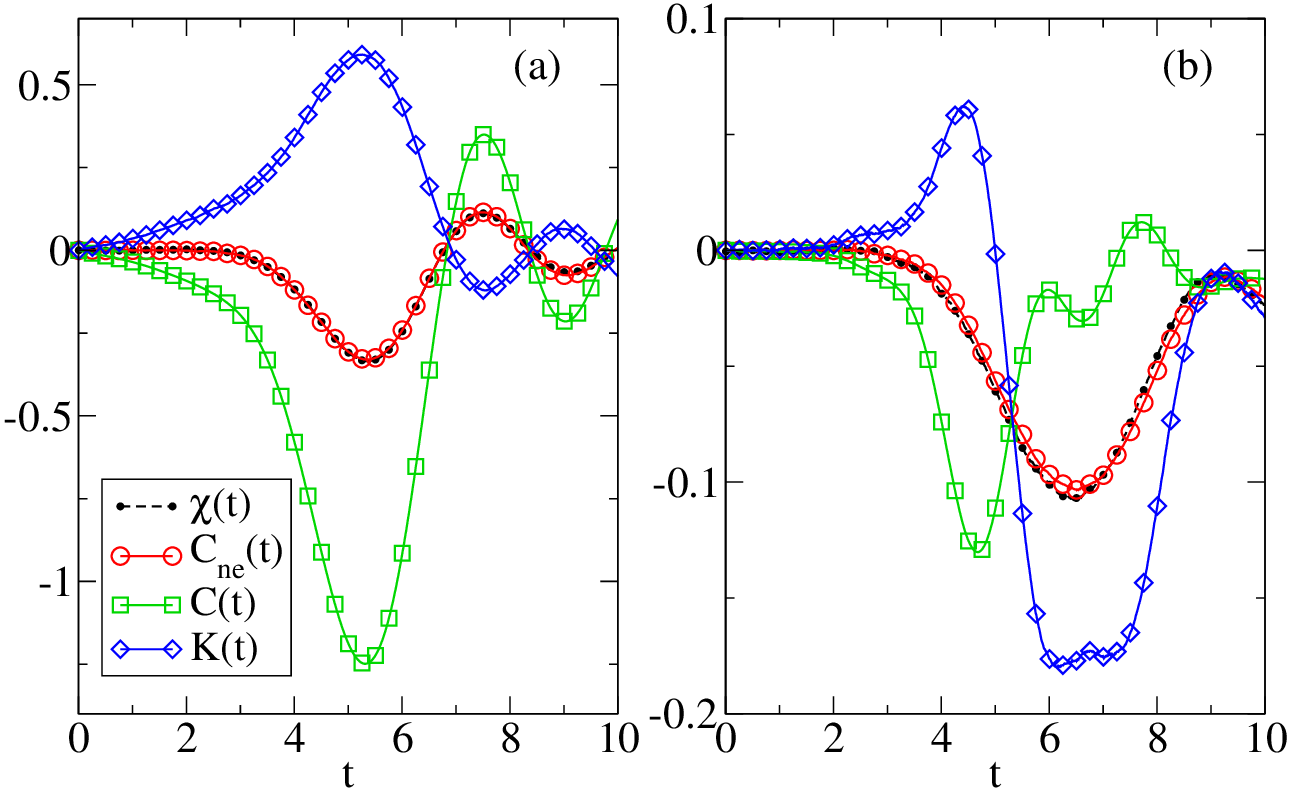}
\end{center}
\caption{
Visualization of the fluctuation-response relation $\chi(t) = C_{\textrm{ne}}(t)$,
for $n=11$ coupled oscillators, with parameters, $D=0.01$, $h=0.03$,
and (a) $T_i=i/10$, (b) $T_i=0.2$ and transient regime as described in the text.
\label{fig:co}}
\end{figure}

Choosing again constant $h=0$ for times $s>0$,
in Fig.~\ref{fig:co}(a) we show the response in a system with $n=11$ oscillators,
with a linear gradient of temperature $T_i = i/10 = 1/\beta_i$, perturbation applied on site $j=1$
and response tested at central site $k=6$.

We finally recall that our formulas work also for transient regimes.
In Fig.~\ref{fig:co}(b) we have the response of a system with constant temperatures $T_i=0.2$,
but where we start from a state out of equilibrium,
by choosing $p_1(0)=10$ and the other momenta equal to zero.

\section{Conclusions}
We have shown that the
nonequilibrium linear response relations for open inertial
dynamics split in two terms.  The first is entropic and makes the
correlation of the observable at time $t>s$ with the excess in
entropy flux at time $s$.  That can involve dissipation over
different reservoirs, each one with its own equilibrium temperature.
The second term depends on more details of the model dynamics and
we have argued it should be understood as an excess in dynamical
activity or generalized escape rate.
So far that activity, which we also called frenesy in linear approximation,
is not understood operationally for real experiments but it is expressed as a
statistical average and correlation function of theoretically
known forces.

 \vspace{1cm}

\noindent {\bf Acknowledgments:} B.W. receives support from  FWO,
Flanders.
M.B. acknowledges financial support from the University of Padua
(Progetto di Ateneo n. CPDA083702) and hospitality from the Institute of Theoretical Physics at the
K.U.Leuven.

 \vspace{1cm}



\bibliographystyle{plain}

\end{document}